\documentclass[letterpaper]{article} % DO NOT CHANGE THIS
\usepackage{aaai2026}
\usepackage{times}  % DO NOT CHANGE THIS
\usepackage{helvet}  % DO NOT CHANGE THIS
\usepackage{courier}  % DO NOT CHANGE THIS
\usepackage[hyphens]{url}  % DO NOT CHANGE THIS
\usepackage{graphicx} % DO NOT CHANGE THIS
\usepackage{natbib}  % DO NOT CHANGE THIS AND DO NOT ADD ANY OPTIONS TO IT
\usepackage{caption} % DO NOT CHANGE THIS AND DO NOT ADD ANY OPTIONS TO IT
\usepackage{booktabs}
\usepackage{amsmath}
\newcommand{\syn}{ZoE}

\title{Folding Pooled Usage Examples into the Tool Index\\Lowers Retrieval of Unused Tools}
\author{Abhijit Chakraborty}
\affiliations{Arizona State University}

\begin{document}
\maketitle

\begin{abstract}
AI agents often find tools by searching their documentation.
Agents sharing a catalog can also pool usage examples: past requests tagged by tool.
What happens to tools no agent has used?
Adding each usage example to the index as an extra entry (folding) helps the other tools but, on two of three catalogs, sharply lowers how often a needed unused tool reaches the top five.
There, using other tools' documentation instead of requests as extra entries left a much smaller loss, though content or style could explain this.
Our rule ZoE (Z-scores over Evidence) scores documentation and usage separately, so tools without usage keep their documentation score.
On all three catalogs, with both embedding models, ZoE finds unused tools more often than a router reserving some of five slots for usage (78.5\% against 68.3\% of queries on one).
With our main embedding model (Jina), the router does better on the other tools on two catalogs, and in post hoc tests adding a constant to used tools' scores showed no detected difference from ZoE on unused tools.
For five simulated agents on one catalog, folding hurt every agent's unused tools; with ZoE no average drop was detected, though some agents lost.
\end{abstract}

\section{Introduction}

An agent that can call tools must first find the right one.
With thousands of tools, a retriever does this: it embeds the request and each tool's documentation, the text that describes what the tool does, and passes the five most similar tools to the agent.
A user asks an agent to fetch every product from a Shopify store and return the full record of one item.
The right tool, a Shopify store scraper, is in the catalog with its documentation, but no agent has used it yet.
Agents that share a catalog can also pool usage examples: past requests, each tagged with the tool that handled it.
Suppose they add each usage example to the retrieval index as an extra entry for its tool; we call this folding.
In our runs on StableToolBench\citep{guo2024stabletoolbench}, one of the three tool catalogs we study, the five tools retrieved for the Shopify request then no longer include the scraper.
With documentation alone, they do.
We chose this query to show the failure, not to estimate how often it happens; the rest of the paper estimates that.

\begin{figure*}[t]
\centering
\includegraphics[width=\textwidth]{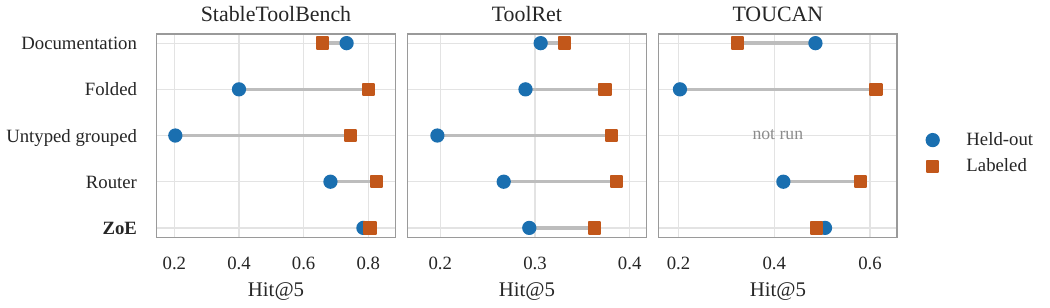}
\caption{Folding usage into the index lowers held-out Hit@5 on StableToolBench and TOUCAN, while \syn{}'s held-out Hit@5 is not detectably below documentation-only. Each row shows one rule's Hit@5 (the share of queries with a correct tool among the five retrieved) on held-out (circle) and labeled (square) test queries, with Jina embeddings, three seeds and settings chosen on development data; note the different x-axis ranges. Documentation-only searches documentation alone. Untyped grouped is folding without the tool's catalog fields (its identifier, category, API names and documentation); it was not run on TOUCAN with Jina.}
\label{fig:main}
\end{figure*}

Pooling past usage is one way a collective of agents shares what its members learn.
Tool-retrieval results are usually reported by benchmark query group \citep{qin2024toolllm,shi2025toolret}, not by whether the needed tool has shared usage.
Yet folded usage can move these two kinds of tools in opposite directions.
We therefore fix a held-out set of tools for which no agent contributes usage.
These are the unused tools of our title, and we report them separately.
We call every other tool labeled, since usage examples may name it.
For a given random draw of the usage pool (a seed, below), a used tool has at least one usage example, and tools without usage are all held-out tools plus the labeled tools that have none in that seed (about a fifth of labeled tools on StableToolBench and most of them on ToolRet; Appendix~A).
Held-out tools keep their documentation in the catalog; held-out means only that no agent has usage for them, not that the embedding model never saw them in pretraining.
The gold tool of a query is the tool the benchmark marks as correct.
Our metric is Hit@5: whether a gold tool is among the five tools passed to the agent.
We report it separately for held-out queries, whose gold tools are all held out, and labeled queries, whose gold tools are all labeled.
It measures retrieval only, not whether the agent then completes the task.
Agents here are simulated: each holds part of one pool of usage records, and no language model runs during retrieval.
We embed text with two general-purpose embedding models, Jina-v2 and Nomic Embed v1.5.
We choose each rule's tunable settings on development tools kept apart from test tools, and we repeat each experiment with three seeds, each with a different random draw of the usage pool.

We compare several rules for combining usage with documentation.
A plain search index holds one entry per tool, its documentation.
Folding adds one entry per usage example, together with its tool's identifier, category, API names and documentation, and scores a tool by its best-matching entry.
A two-index router keeps documentation and usage in separate indexes and reserves some of the five slots for the usage index.
Our rule, \syn{} (Z-scores over Evidence), keeps usage as a second source of evidence about each tool.
Each tool gets one score for its documentation and one for its best-matching usage example.
Each score is standardized across tools for the query (a z-score), and one weight chosen on development data mixes the two.
A tool without usage keeps its own documentation score, although a used tool with a strong usage match can still outrank it.
We also test a constant offset, which adds a constant to the documentation score of every tool with usage.
We test three variants of reciprocal rank fusion (RRF), each combining a tool's rank in the documentation list with its rank in the usage list, and a router that uses its usage slot only when a match passes a threshold.

We ask three questions and find the following.

\paragraph{RQ1: On two catalogs, folding pushes held-out tools out of the top five; on two catalogs, with other tools' documentation as the extra text, the loss is much smaller.}
When a tool is scored by its best entry, extra entries for used tools can only lower the rank of a tool without usage, so the question is how large the loss is.
With Jina on StableToolBench and TOUCAN, folding lowers held-out Hit@5 by about 30 points and raises labeled Hit@5.
On ToolRet both changes are small, possibly because only about a fifth of labeled queries have usage for every gold tool (a post hoc reading).
To check whether the loss depends on what the usage says, a same-length control replaces each usage entry (one usage example in the index) with usage text of the same length, mostly from other tools; we detect no difference in the held-out loss.
On StableToolBench, appending one usage example to each used tool's documentation already causes most of the loss.
A text-type control asks whether the kind of text matters: it fills each usage entry with either the requests or the documentation of one other tool.
On both catalogs where we ran it, documentation causes a much smaller loss than requests; the two replacements also differ in content, so this does not isolate the kind of text.

\paragraph{RQ2: On StableToolBench and TOUCAN, rules that leave tools without usage at their documentation score keep more held-out tools in the top five than folding does.}
Three rules leave tools without usage at their documentation score: the constant offset, the renormalized RRF (a variant that scores such tools by their documentation rank alone) and \syn{}.
With Jina, each keeps held-out Hit@5 at most 2.4 points below documentation-only (point estimates), the baseline that searches documentation alone.
Additive RRF, which gives tools without usage only a down-weighted documentation term, lowers held-out Hit@5 by about 11 points on StableToolBench and 14 on TOUCAN.
\syn{} retrieves more held-out tools than the router on all three catalogs with both embedders.
The router retrieves more labeled tools on TOUCAN with both embedders and on ToolRet with Jina.
\syn{} keeps more of the labeled gain than the offset on StableToolBench with both embedders and on TOUCAN with Jina.

\paragraph{RQ3: For five simulated agents on StableToolBench, folding lowers every agent's held-out Hit@5.}
Each agent retrieves either with its own usage or with all agents' pooled usage.
On StableToolBench, pooling under folding lowers every agent's held-out Hit@5, by 12 to 13 points on average.
Under \syn{} we detect no average held-out change, although in 20 to 47\% of agent-seed pairs the agent loses some held-out Hit@5, and labeled Hit@5 rises.
Separately, on TOUCAN, folding lowers held-out Hit@5 whether 10, 30 or 50\% of tools are held out.

We ask when a collective of agents merely adds cost.
For tool retrieval our answer is concrete: when usage is folded into one index, the cost, in lost Hit@5, falls on tools nobody has used yet, and the combination rule changes how large it is.

\section{Setup}

\paragraph{Catalogs and held-out tools.}
StableToolBench \citep{guo2024stabletoolbench} is a catalog of 3,180 real API tools with test queries in several groups.
Its held-out set is the 172 gold tools of two query groups, G1-tool and G1-category, whose queries use tools or tool categories that the benchmark keeps out of its training data.
Held-out and labeled queries therefore also come from different groups.
The usage pool is the benchmark's instruction pool.
We first remove exact copies of test queries and near-copies (Jina cosine similarity $\geq 0.95$).
We call each request in the pool a source query, and a source query with its gold tools a record. We then drop every record that names a held-out tool as gold or mentions its name, which leaves 90,704 records.
For each seed, we split these records into five parts by primary tool category and sample 5,000 records per part without replacement.
Each sampled record yields one usage entry per gold tool, and removing duplicate query and tool pairs leaves 37,414 to 38,008 usage entries per seed.
A held-out query has every gold tool in the held-out set and a labeled query has none, so a held-out hit always means a held-out tool was retrieved.
A fixed hash of each tool's name also assigns it to a development half, used only to choose settings, or a test half.
Besides the 254 development queries, we exclude the 80 queries with gold tools in both the held-out and labeled sets and the 85 whose gold tools span both halves, which leaves 161 held-out-test and 149 labeled-test queries.
ToolRet \citep{shi2025toolret} is a tool-retrieval benchmark that collects many sources; we exclude its ToolBench, ToolBench-SAM, APIGen and ToolLens sources, which leaves 26,423 tools, 255 held-out-test and 728 labeled-test queries.
TOUCAN \citep{toucan2025} is a corpus of agent trajectories over Model Context Protocol (MCP) servers, which expose tools to agents; our catalog derived from it has 2,159 tools, 769 held-out-test and 2,595 labeled-test queries.
These two have no benchmark groups to hold out, so a fixed hash of the tool name sends 30\% of tools to the held-out set (7,905 and 627 tools), and a second hash assigns the development and test halves.
Neither has a separate usage pool, so a hash of each normalized query sends it either to the usage side or to evaluation (about two in five queries on each).
On ToolRet the usage side holds its own query and gold-tool pairs, not its training pool; on TOUCAN they are question and target-tool pairs from its supervised fine-tuning split, which are requests, not executed trajectories.
We drop usage records that name a held-out tool as gold, mention its name or nearly copy an evaluation query (cosine similarity $\geq 0.95$), and each seed keeps a hash-chosen 80\% of the rest, with at most eight entries per tool.
Each of the three seeds draws a different pool of usage records; the held-out set and the halves stay fixed.
Each source query yields one usage example per gold tool, and these stay together whenever we split the pool.
We call a labeled query fully covered when every gold tool has usage in that seed.
Pooled over seeds, 88\% of labeled-test queries are fully covered on StableToolBench, 21\% on ToolRet and 97\% on TOUCAN, and most covered ToolRet tools have a single usage entry (Appendix~A).

\paragraph{Combination rules.}
For query $q$ and tool $t$, let $a_t$ be the cosine similarity between the embeddings of $q$ and of $t$'s documentation, and $u_t$ the highest cosine similarity between $q$ and any of $t$'s usage entries.
\syn{} scores
\begin{equation*}
s_t = \begin{cases} (1-w)\,z(a_t) + w\,z(u_t) & \text{if } t \text{ has usage,}\\ z(a_t) & \text{otherwise,} \end{cases}
\end{equation*}
where $w \in [0,1]$ is a usage weight chosen on development data (Settings below) and $z(\cdot)$ standardizes a score for the query: it subtracts the mean and divides by the standard deviation, taken over the whole catalog for $a_t$ and over the tools that have usage in that seed for $u_t$.
The design follows from how folding ranks tools.
Under folding, a used tool scores as its best entry, so every usage entry is another chance to match the query, while a tool without usage has only its documentation.
\syn{} gives every tool one documentation score on the same scale and adds usage as a separate, standardized term.
A tool without usage keeps its standardized documentation score, and a used tool's entries act through one weighted usage term instead of each competing as a separate entry.
A used tool can still outrank it through its usage term, so \syn{} does not guarantee the rank of a tool without usage.
RQ1 and RQ2 test parts of this reasoning; they do not isolate it.
Folding ranks all index entries and walks down the list until five distinct tools appear, so adding entries for used tools can never raise the rank of a tool without usage.
Each folded entry is a short text, with a field name before each part, that gives the tool's identifier, a tool description made of its catalog category, up to three API names and its documentation, and then the usage request; we call these added parts the tool's catalog fields.
When a usage record carries a source type (where the record came from), a one-line note of that type is also added.
The untyped grouped variant scores each tool by its best entry among its documentation and its raw usage requests, without catalog fields or field names (Raw, all in Appendix~J).
The other rules use the raw usage requests: \syn{}'s $u_t$, the router's usage index and the fusion rules' usage list are all built from them.
The router takes $d$ tools from a documentation index and fills the remaining slots from a usage index.
The constant offset adds one development-tuned value to the standardized documentation score of every tool with usage.
Reciprocal rank fusion (RRF) scores a tool from its rank in the documentation list and its rank in the usage list, the list of tools with usage ordered by their best-matching usage request \citep{cormack2009rrf}.
Additive RRF sums the two weighted reciprocal ranks, so a tool without usage gets only the down-weighted documentation term.
The renormalized variant scores such a tool by its documentation rank alone, without the down-weighting, and the best-field variant takes the larger weighted term.
The thresholded router gives one slot to the best usage match only when its cosine similarity passes a threshold, set so that 5\% of held-out development queries pass it (a 5\% false-positive rate).

\paragraph{Settings and statistics.}
All rules use Jina-v2 embeddings \citep{gunther2023jina} and return five distinct tools; we repeat the main comparisons with Nomic Embed v1.5 \citep{nussbaum2024nomic} (Appendix~G).
\syn{}'s weight $w$ and the router's number of documentation slots $d$ were selected on development tools and fixed before the final scoring run reported here; we call these the frozen settings.
The test sets had been seen during earlier development, so frozen means that the settings were not changed after this run, not that the test sets were unseen.
The frozen settings are $w{=}0.5$, $d{=}4$ on StableToolBench; $w{=}0.9$, $d{=}4$ on ToolRet; and $w{=}0.9$, $d{=}3$ on TOUCAN.
The offset, the fusion weights and rank constants, and the threshold were also selected on development data, but after the test sets had been seen in earlier versions of this work (Appendices I and L).
Intervals are paired 95\% percentile bootstrap intervals with 10,000 draws.
They resample clusters of queries that share a first gold tool (in alphabetical order) and keep all three seeds of a cluster together.
An interval that includes zero is reported as no difference detected, not as equivalence.
Appendix~F and the tables give the values behind the main figures.
On StableToolBench, \syn{} keeps its held-out lead over the router when tools are renamed to opaque names (Appendix~D).
Only two analyses were planned before the final scoring run: \syn{} against the router at five candidates, and the criterion for replication with Nomic (\syn{}'s held-out lead over the router on all three catalogs and the router's labeled lead on ToolRet and TOUCAN).
Every other comparison is descriptive (Appendix~B).

\section{RQ1: What Folding Costs Unused Tools}

\paragraph{With Jina, folding sharply lowers held-out Hit@5 on two of three catalogs.}
Figure~\ref{fig:main} shows each catalog.
Folding raises labeled-test Hit@5 by 14 points on StableToolBench and 29 on TOUCAN.
It lowers held-out Hit@5 from 0.733 to 0.400 and from 0.486 to 0.203.
On ToolRet both changes are small: held-out Hit@5 falls from 0.306 to 0.290 ($-1.57$ points, [$-3.19$, $-0.38$]), a small but detected loss, possibly because usage covers few labeled queries: on fully covered ones, folding raises labeled Hit@5 from 0.136 to 0.385.
This explanation is post hoc.
With Nomic, folding again lowers held-out Hit@5 (by 16.15, 1.83 and 22.63 points; every interval below zero).
The untyped grouped variant loses even more held-out Hit@5 on the two catalogs where we ran it with Jina.

\paragraph{The held-out loss remains with usage text mostly about other tools, and on StableToolBench one appended example causes most of it.}
Table~\ref{tab:rq1} collects the StableToolBench controls.
To test whether the loss depends on what the usage says, we keep every tool's catalog fields and number of usage entries.
In this same-length control, we replace each entry with text from other tools' usage of exactly the same tokenizer length (Appendix~E, Figure~\ref{fig:length}, Table~\ref{tab:length}).
Held-out Hit@5 differs from the real-usage value by $+2.28$ points ([$-0.20$, 5.29]) on StableToolBench and $-0.87$ ([$-2.06$, 0.24]) on TOUCAN, so no difference is detected.
Labeled Hit@5 falls by 33 and 27 points.
With Nomic on StableToolBench the pattern repeats (held-out $+1.86$, [$-1.24$, 5.10]; labeled $-12.75$).
So the labeled gain needs text about the right tool; we detect no such need for the held-out loss.

On StableToolBench, one appended usage example per tool causes most of the loss; more separate entries add to it (one-entry folding 0.553 against folding 0.400, descriptive), while appending all usage instead of one example gives about the same value (0.424 against 0.429, point estimates).
Appending one usage example to each used tool's documentation, so that every tool keeps exactly one entry, lowers held-out Hit@5 from 0.733 to 0.429 ($-30.43$, [$-37.50$, $-23.53$]).
Appending all usage instead shows no detected difference from folding ($+2.48$, [$-1.39$, 6.58]).
One-entry folding, which keeps one separate usage entry per used tool, gives a smaller loss, to 0.553 (Appendix~I).
Giving every tool the same number of entries does not remove the loss either.
With one documentation entry and one request per tool (a real usage example for used tools, a request that a language model generated from the documentation for the rest), held-out Hit@5 is 0.472 ($-26.09$, [$-33.74$, $-18.43$]).
The generated requests raise it by 11.39 points ([6.67, 16.86]) over an index with one documentation entry per tool and one raw usage request per used tool (Raw, one in Appendix~J), without the generated requests, so query-like text for tools without usage helps them but does not close the gap (Appendix~J).

\begin{table}[t]
\centering
\small
\setlength{\tabcolsep}{4pt}
\caption{StableToolBench controls: Hit@5 on held-out-test (H) and labeled-test (L) queries, Jina, three seeds. Rows below Folded are index variants described in the text; ``one request per tool'' is a real usage example for used tools and a generated request for the rest.}
\label{tab:rq1}
\begin{tabular}{lcc}
\toprule
Index & H & L \\
\midrule
Documentation only & .733 & .658 \\
Folded (real usage) & .400 & .801 \\
Other tools' usage, same length & .422 & .474 \\
One example appended to documentation & .429 & .671 \\
Documentation + one request per tool & .472 & .530 \\
Other tools' requests (text-type control) & .424 & .510 \\
Other tools' documentation (text-type control) & .640 & .653 \\
\bottomrule
\end{tabular}
\end{table}

\begin{figure}[t]
\centering
\includegraphics[width=\columnwidth]{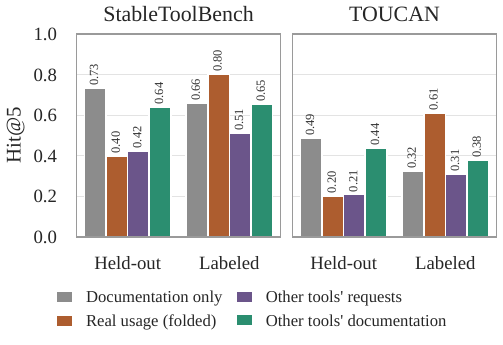}
\caption{Text-type control: other tools' requests used as usage give held-out Hit@5 2.5 points above folding's on StableToolBench and not detectably different on TOUCAN, and other tools' documentation matched to each entry's token length gives a much smaller loss. Both replacements come from the same donor tool and match each entry's token length; Hit@5, Jina, three seeds.}
\label{fig:register}
\end{figure}

\paragraph{On StableToolBench and TOUCAN with Jina, replacing usage with other tools' documentation leaves a much smaller loss than replacing it with their requests.}
The text-type control asks whether the kind of text matters: requests or documentation.
For each usage entry we take one donor tool and replace the entry either with the donor's request text or with the donor's documentation, matched to the entry's exact token length.
The entry keeps its own tool, catalog fields and position (Appendix~L).
Figure~\ref{fig:register} shows the result.
Other tools' requests give held-out Hit@5 of 0.424 on StableToolBench and 0.209 on TOUCAN.
That is 2.48 points above folding on StableToolBench ([0.21, 5.13]), and on TOUCAN no difference from folding is detected.
Other tools' documentation gives 0.640 and 0.439, which is 21.53 ([15.04, 28.29]) and 22.97 ([12.54, 32.88]) points above requests.
Many replacements repeat their text to reach the length: about half of the request replacements on both catalogs, and 62\% of the documentation replacements on StableToolBench but 7\% on TOUCAN.
On entries where neither replacement needed repetition, the gap is smaller but remains: 12.42 points ([7.25, 18.13]) on StableToolBench and 16.86 ([8.96, 24.45]) on TOUCAN.
That filter also changes which tools keep usage, so it is a sensitivity check, not a clean test.
Documentation text is not harmless: on StableToolBench it still lowers held-out Hit@5 by 9.32 points against documentation-only ([5.32, 13.68]).
We observe a larger loss with request replacements than with documentation replacements, and a smaller but detected loss with documentation replacements than with no added text.
One interpretation, which this control does not test, is that request-shaped text moves a used tool's entries toward requests in general.
The control does not isolate the kind of text, because the two replacements also differ in content.
In each benchmark the usage requests and the test queries also come from the same sources, so a match to their shared writing style is an equally plausible reading that we did not test.

\section{RQ2: Which Rules Keep Unused Tools in the Top Five}

\begin{figure*}[t]
\centering
\includegraphics[width=\textwidth]{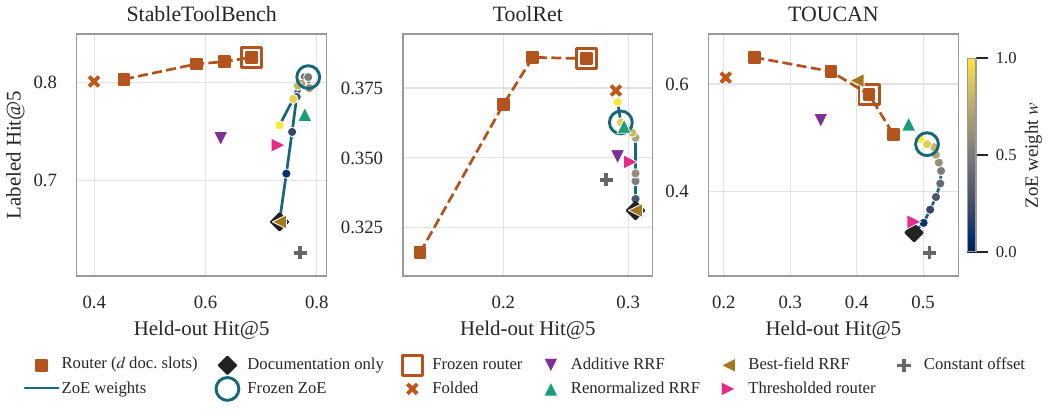}
\caption{On StableToolBench and TOUCAN, rules that keep tools without usage at their documentation score sit at the right of each panel (high held-out Hit@5), though the constant offset loses labeled Hit@5 on StableToolBench and TOUCAN (point estimates); folding, additive RRF and, on TOUCAN, best-field RRF lose held-out Hit@5. Held-out-test against labeled-test Hit@5, Jina, three seeds. The \syn{} curve runs from documentation only ($w{=}0$, diamond) to $w{=}1$, where $w$ is \syn{}'s usage weight; squares are router settings with $d$ documentation slots out of five. Circles mark the settings frozen on development data before test scoring; all other points were selected or plotted after the frozen run. Table~\ref{tab:sweep} gives the values. On StableToolBench and ToolRet, best-field RRF returns the same lists as documentation-only.}
\label{fig:rules}
\end{figure*}

\begin{table*}[t]
\centering
\small
\setlength{\tabcolsep}{5pt}
\caption{Main combination rules on three catalogs: Hit@5 on held-out-test (H) and labeled-test (L) queries, Jina, three seeds. Rules in the lower block leave tools without usage at their documentation score. Only the router and \syn{} settings were frozen before test scoring; documentation and folding have no settings.}
\label{tab:rules}
\begin{tabular}{lcccccc}
\toprule
 & \multicolumn{2}{c}{StableToolBench} & \multicolumn{2}{c}{ToolRet} & \multicolumn{2}{c}{TOUCAN} \\
Rule & H & L & H & L & H & L \\
\midrule
Documentation only & .733 & .658 & .306 & .331 & .486 & .323 \\
Folded & .400 & .801 & .290 & .374 & .203 & .612 \\
Additive RRF & .627 & .743 & .292 & .350 & .346 & .532 \\
Router & .683 & .826 & .267 & .386 & .419 & .580 \\
Thresholded router & .731 & .736 & .302 & .348 & .486 & .343 \\
Best-field RRF & .733 & .658 & .306 & .331 & .401 & .606 \\
\midrule
Constant offset & .770 & .626 & .282 & .342 & .509 & .286 \\
Renormalized RRF & .778 & .767 & .297 & .361 & .478 & .525 \\
\syn{} & .785 & .805 & .294 & .363 & .506 & .488 \\
\bottomrule
\end{tabular}
\end{table*}

\paragraph{\syn{} retrieves more held-out tools than the router; the router retrieves more labeled tools on two catalogs.}
Table~\ref{tab:rules} lists every rule.
Against documentation alone, \syn{} changes held-out Hit@5 by $+5.18$ points on StableToolBench ([2.31, 8.39]), $-1.18$ on ToolRet ([$-2.85$, 0.24]) and $+1.95$ on TOUCAN ([$-0.39$, 4.52]).
It raises labeled-test Hit@5 by 14.77 ([5.51, 22.55]), 3.16 ([0.60, 6.18]) and 16.49 ([7.49, 25.31]) points.
Against the router, \syn{} is higher on held-out queries on all three catalogs: by $+10.14$ ([5.68, 15.05]), $+2.75$ ([0.67, 5.06]) and $+8.71$ ([5.15, 12.39]) points.
The router is higher on labeled-test queries on ToolRet, by 2.29 points ([0.04, 4.85]), and on TOUCAN, by 9.24 ([5.85, 12.78]).
On StableToolBench no labeled difference is detected ($-2.01$, [$-5.81$, 1.46] for \syn{} minus router).
On ToolRet the router's lead grows to 17.15 points ([8.73, 27.45]) on fully covered queries.
When we sweep both settings after the test sets had been seen (Figure~\ref{fig:rules}, Table~\ref{tab:sweep}), every \syn{} weight has higher held-out Hit@5 than every tested router setting ($d$ from 1 to 4) on all three catalogs, and some router settings have higher labeled Hit@5 than every weight.
With 10 and 30 candidates, the held-out lead narrows on StableToolBench and ToolRet and stays near 9 points on TOUCAN (Appendix~C).

\paragraph{With Jina, other rules that keep the documentation score for tools without usage differ from \syn{} by at most 2.8 held-out points (point estimates).}
Appendices I and L give every rule's settings and paired intervals; among these rules, only the renormalized RRF on TOUCAN differs detectably from \syn{} on held-out queries.
Additive RRF, where a tool without usage gets no usage term, lowers held-out Hit@5 against documentation by 10.56 ([6.25, 15.37]) points on StableToolBench and 14.04 ([7.91, 20.55]) on TOUCAN.
The renormalized RRF, which scores a tool without usage by its documentation rank alone, is within 0.62 held-out points of \syn{} on StableToolBench (no difference detected, [$-0.22$, 1.57] for \syn{} minus RRF) and 2.77 points below it on TOUCAN ([0.61, 4.98]).
On labeled queries \syn{} is 3.80 points higher on StableToolBench ([0.43, 7.60]), and on TOUCAN no difference is detected.
On ToolRet, where held-out effects are small, neither any fusion rule nor the thresholded router differs detectably from \syn{} on held-out queries.
The constant offset shows no detected held-out difference from \syn{} in five of six embedder and catalog settings.
It is 10 to 20 points lower on labeled queries on StableToolBench with both embedders and on TOUCAN with Jina (Appendix~I).
\syn{} has higher labeled Hit@5 than the thresholded router on StableToolBench and TOUCAN (by 6.94 and 14.46 points) and higher held-out Hit@5 on StableToolBench (5.38, [2.43, 8.71]).
We also built a Re-Invoke-inspired index that adds one request, generated from documentation by Llama-3.1-8B-Instruct, to every tool as a separate entry \citep{chen2024reinvoke}.
\syn{} is higher than this index on StableToolBench by 12.01 held-out ([6.40, 18.16]) and 14.77 labeled points ([4.51, 25.31]; Appendix~J).
Our interpretation is that the score a tool without usage receives matters more than the exact form of the combination.
\syn{}'s particular choice, two standardized scores with one weight, is not better than every alternative.
We detect no held-out difference from the renormalized RRF on two catalogs, and the router keeps more labeled gain on two.

\paragraph{With another tool's usage, \syn{}'s held-out Hit@5 does not fall, and its labeled Hit@5 does.}
In a permuted-usage control, each tool's usage is replaced by another tool's usage, without matching length per entry, and \syn{} keeps its frozen weight.
Held-out Hit@5 stays the same or rises in all six embedder and catalog settings, while labeled Hit@5 falls by 6.2 to 54.6 points (Appendix~I, Table~\ref{tab:othertool}).
Together with the constant-offset result, this is consistent with two separate parts: held-out Hit@5 depends on how tools without usage are scored against the rest, and the labeled gain depends on usage about the right tool.
The controls do not isolate which part of the scoring rule keeps held-out tools in the top five.

\paragraph{A second embedder keeps the held-out result and weakens two labeled results.}
We reran every main rule with Nomic Embed v1.5 and reselected \syn{}'s usage weight $w$ and the router's number of documentation slots $d$ on development tools (StableToolBench $w{=}0.5$, $d{=}4$; ToolRet $w{=}1.0$, $d{=}4$; TOUCAN $w{=}0.7$, $d{=}3$).
\syn{} again beats the router on held-out queries on all three catalogs: by 7.66 points ([2.63, 13.14]), 3.14 ([1.25, 5.34]) and 11.05 ([7.55, 14.56]).
The router again beats \syn{} on TOUCAN labeled queries, by 14.63 points ([11.72, 17.54]).
On ToolRet the router's labeled lead is 0.92 points with an interval that includes zero ([$-0.79$, 2.70]); on fully covered ToolRet queries it still leads by 10.47 points ([3.43, 18.48]).
On TOUCAN, \syn{}'s labeled gain over documentation shrinks to 1.28 points ([$-4.63$, 6.86]).
By the criterion we set before the run (held-out lead on all three catalogs, router labeled lead on ToolRet and TOUCAN), replication is therefore not established.

\paragraph{On StableToolBench, the better rule depends on the workload.}
On StableToolBench, \syn{} is ahead of the router when fewer than 83\% of queries go to labeled tools (95\% interval of this break-even share: 59 to 100\%).
Above 59\%, the data do not decide.
Held-out and labeled queries also come from different benchmark groups, so the estimate mixes tool status with query source (Appendix~H).

\section{RQ3: What Pooling Costs Each Agent}

\begin{figure}[t]
\centering
\includegraphics[width=\columnwidth]{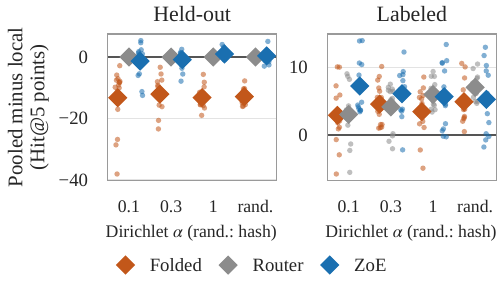}
\caption{Pooling lowers every agent's held-out Hit@5 under folding; under \syn{} no average change is detected. Change in each agent's Hit@5 when it retrieves with all agents' usage instead of its own (dots: 5 agents $\times$ 3 seeds; diamonds: means), StableToolBench, Jina. The router's held-out change is zero because held-out tools have no usage in any pool. Pooled minus local: Hit@5 with all agents' usage minus Hit@5 with the agent's own. $\alpha$: Dirichlet concentration, where smaller values give agents more different tool mixes; hash: hash split.}
\label{fig:m4}
\end{figure}

\begin{figure}[t]
\centering
\includegraphics[width=\columnwidth]{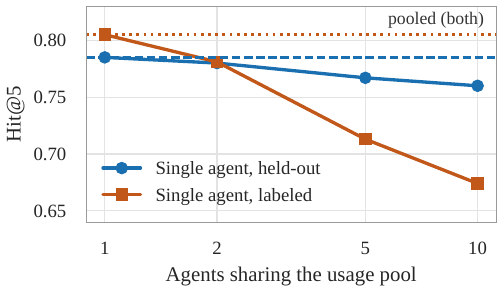}
\caption{Hash split with similar tool mixes: a single agent's \syn{} Hit@5 drops as the usage pool is split among more agents, most on labeled queries; pooling recovers it. Horizontal lines, marked ``pooled (both)'', are pooled Hit@5 (dashed: held-out; dotted: labeled). StableToolBench, averaged over agents and three seeds.}
\label{fig:agents}
\end{figure}

\paragraph{Under folding, pooling lowers every agent's held-out Hit@5.}
We split StableToolBench's usage pool among five simulated agents by tool category.
Each source query goes to one agent.
For each set of tool categories, the share of its records that each of the five agents receives is drawn at random from a Dirichlet distribution with concentration $\alpha$ (0.1, 0.3 or 1), following the per-class partition of \citet{wang2020fedma}; Dirichlet splits are a common way to simulate uneven client data in federated learning \citep{hsu2019noniid}.
A small $\alpha$ makes the split uneven: at $\alpha{=}0.1$ a category's records tend to go mostly to one agent, so agents specialize in different tools, while at $\alpha{=}1$ their tool mixes overlap more.
In a fourth split, a hash of each source query picks its agent.
We weight each agent's test queries by the category mix of its own records, so its test workload follows its own tool mix (Appendix~K).
Each agent retrieves either with its own usage or with all agents' usage, at the frozen settings.
Under folding, pooling lowers held-out Hit@5 for every agent and seed in all four splits (Figure~\ref{fig:m4}), by 12.10 points on average at $\alpha{=}0.3$ ([$-19.04$, $-5.93$]).
The direction follows from best-entry scoring; the size is what we measure.
Single-agent extremes are exploratory, because some workloads are small: the smallest has a Kish effective sample size (the number of equally weighted queries that would give the same variance) of 5.9 held-out queries.
Under \syn{}, the average held-out change is $-0.86$ at $\alpha{=}0.3$ ([$-4.58$, 2.05]), and it lies between $-1.32$ and $+0.97$ in all splits, each interval including zero.
The average hides losers: in 20 to 47\% of agent-seed pairs, the agent loses held-out Hit@5.
Taking the worst agent in each seed and averaging over seeds, the loss is 9.93 points at $\alpha{=}0.1$, and one agent loses 12.5 points in one seed.
\syn{}'s labeled gain from pooling is 5.26 to 7.25 points, each interval above zero.
A pooled list equals that of one agent holding the same records, so these gains come from access to more records; we test no mechanism for sharing records across separate parties.
We also split the pool by hash among 1, 2, 5 or 10 agents, so that every agent gets a similar tool mix, and score all test queries with equal weight (Figure~\ref{fig:agents}; Appendix~K compares it with the hash split above).
A single agent's \syn{} Hit@5 then falls as the pool is divided, from 0.805 to 0.674 on labeled queries at ten agents, and pooling recovers it.
At ten agents, pooling beats the average single agent by 2.48 held-out points ([0.64, 4.51]) and 13.18 labeled points ([6.04, 18.70]).

\begin{figure}[t]
\centering
\includegraphics[width=\columnwidth]{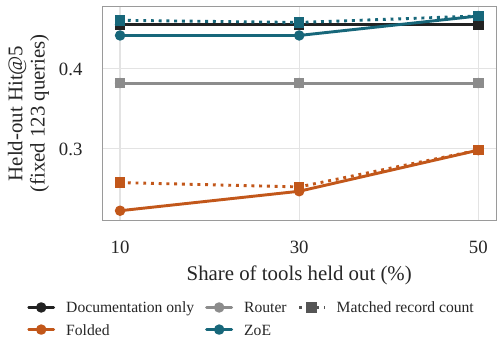}
\caption{Folding lowers held-out Hit@5 at every held-out share we tried, and \syn{} is between 1.4 points below and 1.1 points above documentation-only (point estimates; no paired interval). TOUCAN, the 123 queries held out at all three shares (21 tool clusters); solid lines use all remaining usage, dotted lines subsample it to the same record count at every share; the router is at its frozen setting.}
\label{fig:share}
\end{figure}

\paragraph{The cost to held-out tools appears at every held-out share we tried.}
On TOUCAN we held out 10, 30 or 50\% of tools with nested hash sets, so the 10\% set lies inside the other two, and dropped the usage records of newly held-out tools (Appendix~M).
On the 123 queries held out at all three shares, folding lowers held-out Hit@5 against documentation by 23.31 ([13.82, 39.93]), 20.87 ([12.15, 36.82]) and 15.72 ([8.55, 28.02]) points at 10, 30 and 50\% (Figure~\ref{fig:share}).
\syn{} is 16.80 to 21.95 points above folding at every share.
Subsampling the usage to the same number of records at every share changes these values by at most 3.6 points (point estimates; Table~\ref{tab:share}).
With 21 tool clusters we do not test a trend across shares.

\section{What This Means for Agent Collectives}

Three practical points follow from these results, each within the scope of our tests.
First, an evaluation of pooled agent experience should report tools or tasks that no member has covered separately from the average.
In our agent study, pooling raised labeled Hit@5 in every split under \syn{}.
Under folding it raised labeled Hit@5 on average and lowered every agent's held-out Hit@5.
Averaged over all queries at the observed mix of 161 held-out and 149 labeled test queries, folding would show a loss of about 4 points at $\alpha{=}0.3$ (4 to 5.5 across splits), much smaller than its 12-point loss on held-out queries, and \syn{} would show a gain.
Second, folding shared records into one index has a cost that falls on tools no member has used, large on StableToolBench and TOUCAN, and on TOUCAN that cost appeared whether 10, 30 or 50\% of tools were held out.
Third, the combination rule is a design choice with a workload trade-off.
With Jina, rules that keep tools without usage at their documentation score had held-out Hit@5 at most 2.4 points below documentation-only (point estimates), and a router that reserves slots for usage retrieved more labeled tools on two catalogs.
On TOUCAN, where 97\% of labeled queries are fully covered, \syn{} keeps about 57\% of folding's labeled gain with Jina and almost none with Nomic, so either choice has a cost.
We estimated where this trade-off crosses only on StableToolBench (RQ2, Appendix~H), so the choice should follow the expected share of requests to unused tools.

The collective question our results answer is who pays when members' experience is merged into one index.
In our tests, it is the members whose requests go to tools nobody has used.

\section{Related Work}

Tool retrievers select from documentation \citep{qin2024toolllm}, and off-the-shelf retrievers find tool retrieval hard \citep{shi2025toolret}.
Expanding documents with predicted queries improves retrieval \citep{nogueira2019doc2query}, and filtering out generated queries that do not match their document improves it further \citep{gospodinov2023doc2queryminus}.
Re-Invoke expands every tool with generated queries \citep{chen2024reinvoke}.
Tool2Vec represents each tool by the average embedding of the queries that used it \citep{moon2024tool2vec}, and Tool-Embed is trained on tool documentation expanded by a language model \citep{lu2026underdocumented}.
These works measure what expansion does to the tools that receive it; \citet{lu2026underdocumented} also find that the example-usage field adds the least, often lowering accuracy, and that expanded documents receive lower similarity scores, more for irrelevant tools than for relevant ones.
We study the tools left without expansion when they compete with expanded ones.
On the two catalogs we tested with Jina, their loss is much smaller when the expanded tools carry documentation text than when they carry request text, though content and benchmark writing style could also explain this.
Combining evidence from several fields is an old idea in retrieval: BM25F weights fields inside one score \citep{robertson2004bm25f}, and mFAR scores fields separately and learns how to mix them \citep{li2025mfar}; MFTR (multi-field tool retrieval) learns one weight per tool field, including a field of example user intents \citep{tang2026mftr}.
Our contribution is the measurement and the comparison of how such rules treat tools without usage, not a new scoring form.
Multi-agent systems share memory or messages through a common pool \citep{hong2024metagpt,gao2024memshare}, and agents reuse their own experience as distilled insights \citep{zhao2024expel} or induced workflows \citep{wang2025awm}.
These works evaluate the agent that receives the memory, while we look at retrieval for the tools that the memory does not cover.
Federated agent learning shares free-text knowledge or in-context examples across clients \citep{wu2024fical,wang2025fedicl} or learns language-model routers from decentralized evaluations \citep{askin2026federate}.

\section{Limitations}

All results are retrieval metrics, mainly Hit@5, with two general-purpose embedders.
We have not tested task success, a retriever trained for tools such as Tool-Embed, a cross-encoder reranker (a model that rescores each query and tool pair) or the full Re-Invoke.
Our audit of Nomic's training data is incomplete, so we do not claim that it never saw these benchmarks.
Agents are simulated partitions of one usage pool with reweighted workloads, and no privacy mechanism is applied.
We report held-out tools separately but not labeled tools that lack usage in a seed, which stay inside the labeled results.
Held-out tools on StableToolBench are a benchmark-chosen subset, and on ToolRet and TOUCAN a hash sample; neither models which tools are new in a real deployment.
The test sets were used in earlier versions of this work, and only \syn{} against the router and the Nomic replication criterion were fixed before scoring.
The offset, fusion and threshold settings were chosen on development data afterwards, which may favor those baselines.
We removed test-query copies from the usage pool only at Jina cosine similarity $\geq 0.95$, so labeled gains may include some leakage of paraphrased test queries.
Held-out tools have no usage entries that could leak, but leaked entries of labeled tools still compete in held-out rankings and could move held-out Hit@5 in either direction.
The equal-entry controls (Appendix~J) and the agent study ran on StableToolBench, the text-type control and fusion rules on two or three catalogs, and the held-out-share sweep on TOUCAN with 21 tool clusters, all with Jina only.
On ToolRet, \syn{} does not improve held-out Hit@5 over documentation alone.

\section{Conclusion}

Shared usage helps the tools it describes. When folded into one index, it lowers how often tools nobody has used reach the five candidates, by about 30 points on two of three catalogs with Jina.
On StableToolBench and TOUCAN with Jina, with documentation text in place of requests, the loss is much smaller, though we could not separate the kind of text from its content or from benchmark style.
When a large share of requests goes to such tools, rules that leave tools without usage at their documentation score, such as \syn{} or a renormalized rank fusion, lost far less than folding on StableToolBench and TOUCAN with Jina; apart from \syn{} against the router, these comparisons are descriptive.
The router retrieved more labeled tools on TOUCAN with both embedders and on ToolRet with Jina.
Reports on pooled agent experience should also give the held-out number.

\bibliography{refs}
\clearpage
\appendix

\section{Appendix A: Splits and query accounting}

\paragraph{StableToolBench.}
The held-out set is the union of normalized gold tools in the G1-tool (95 tools) and G1-category (77 tools) groups.
The usage pool excludes any record that names a held-out tool as gold or mentions its name; the mention check covers the query, gold tool names and attached tool, API and category fields after lowercasing and mapping underscores and hyphens to spaces.
A fixed hash of each tool's name sends it to the development or test half with equal probability, so the split does not depend on the tool's content or results.
Table~\ref{tab:exclusions} accounts for all 729 queries.
The three seeds change which usage records are present (they cover 2,430, 2,436 and 2,422 tools), not the held-out set or the half of any tool.

\begin{table}[h]
\centering
\small
\caption{StableToolBench queries left out of the test results, counted once per query.}
\label{tab:exclusions}
\begin{tabular}{lr}
\toprule
Population & Queries \\
\midrule
Held-out, development (tuning only) & 135 \\
Labeled, development (tuning only) & 119 \\
Held-out, gold tools span halves & 13 \\
Labeled, gold tools span halves & 72 \\
Gold tools both held out and labeled & 80 \\
\midrule
Total left out & 419 \\
Test queries (161 held-out, 149 labeled) & 310 \\
\bottomrule
\end{tabular}
\end{table}

\paragraph{ToolRet and TOUCAN.}
A fixed hash of each tool's name holds out 30\% of tools, giving 7,905 of 26,423 on ToolRet and 627 of 2,159 on TOUCAN. A second, independent hash sends each tool to the development or test bucket with equal probability.
After source exclusions and exact-document alias merging, ToolRet has 26,423 tools from 31 source datasets.
A query audit against the ToolBench instruction pool found no exact matches and removed 4 queries at Jina cosine $\geq 0.95$; this does not rule out every semantic overlap.

\paragraph{What held-out stands for.}
Held-out tools are tools for which no agent contributed usage; their documentation stays in the catalog and we do not control what the embedders saw in pretraining.
On StableToolBench they are the benchmark's G1-tool and G1-category gold tools, a subset chosen by the benchmark, not at random.
On ToolRet and TOUCAN they are a 30\% hash sample of the catalog, which is random with respect to tool content but not a model of which tools are new in a real deployment.
\paragraph{Usage caps.}
ToolRet and TOUCAN keep a hash-chosen 80\% of usage queries and at most eight usage entries per tool in each seed, while the StableToolBench pools have no per-tool cap (at most 380, 445 and 343 entries per tool across seeds).
Most covered ToolRet tools have a single entry (773 to 792 of 863 to 885 covered tools per seed, from 1,023 to 1,049 usage entries).

\paragraph{Usage coverage of labeled queries.}
Table~\ref{tab:coverage} counts labeled test queries once per seed (a query scored in three seeds counts three times) and the share that is fully covered.
ToolRet's covered-only comparisons therefore rest on about a fifth of its labeled rows.

\begin{table}[h]
\centering
\small
\caption{Labeled-test rows pooled over three seeds, and rows whose every gold tool has usage.}
\label{tab:coverage}
\setlength{\tabcolsep}{3pt}
\begin{tabular}{lrrr}
\toprule
Catalog & Labeled & All gold covered & Share \\
\midrule
StableToolBench & 447 & 394 & 88\% \\
ToolRet & 2,184 & 449 & 21\% \\
TOUCAN & 7,785 & 7,533 & 97\% \\
\bottomrule
\end{tabular}
\end{table}

\section{Appendix B: Models, metric and statistics}

Jina-v2 is jina-embeddings-v2-base-en and Nomic Embed v1.5 is nomic-embed-text-v1.5; the three seeds are 42, 123 and 456.
\syn{}'s weight $w$ is selected from $\{0, 0.1, \ldots, 1\}$ and the router's $d$ from $\{1, 2, 3, 4\}$; Appendices I and L give the offset and fusion grids and the threshold rule.
Hit@5 gives credit when any gold tool is among five distinct candidates; on single-gold queries it equals Recall@5. Hit@$k$ is the same with $k$ candidates.
Multi-gold queries are rare among held-out queries (4 of 161 on StableToolBench, 4 of 255 on ToolRet) and more common among labeled ones (62 of 149 and 95 of 728).
Folded and untyped grouped indexes rank index entries and walk down the list until five distinct tools appear, up to 200 entries; every list reached five tools before the cap, so the cap never changed a result.
Intervals resample first-gold-tool clusters with all seeds and queries of a cluster kept together.
Multi-gold queries can share non-first tools across clusters, so this convention does not capture every dependence.
Intervals are conditional on the frozen settings and do not include selection uncertainty, and we do not adjust them for multiple comparisons.
Two analyses were planned before the final scoring run: \syn{} against the router on held-out and labeled test queries at five candidates, and the Nomic replication criterion; all other comparisons are descriptive.
Because the test sets had been examined in earlier development, even the planned analyses are not an untouched confirmatory test, and their intervals are conditional on the selected settings.
The number of first-gold-tool clusters behind the five-candidate intervals is 92 (held-out), 109 (labeled) and 98 (fully covered) on StableToolBench, 220, 577 and 81 on ToolRet, and 89, 234 and 190 on TOUCAN.

\section{Appendix C: Retrieval budget}

\begin{figure}[!htbp]
\centering
\includegraphics[width=\columnwidth]{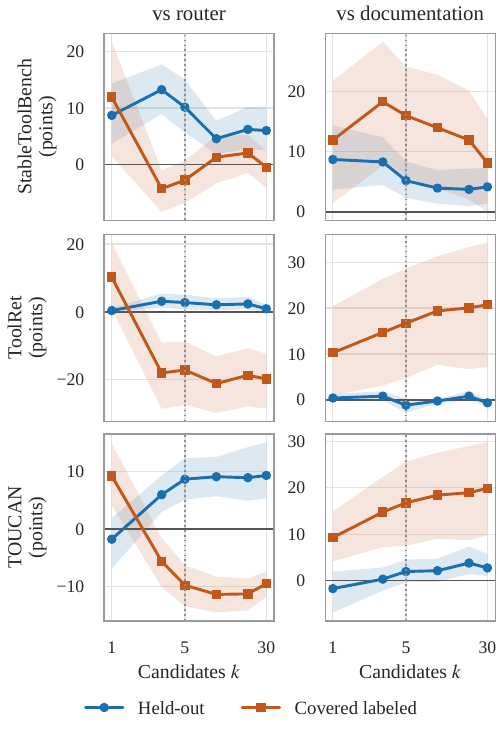}
\caption{\syn{}'s held-out lead over the router narrows with more candidates on StableToolBench and ToolRet but not on TOUCAN, and its covered-query gain over documentation stays positive at 5, 10 and 30 candidates (the StableToolBench interval at 30 touches zero). Paired differences in Hit@$k$ points with tool-clustered 95\% intervals (bands); covered labeled means every gold tool has usage. The router keeps its $k{=}5$ share of slots for usage (one in five on StableToolBench and ToolRet, two in five on TOUCAN), rounded to whole slots, so at $k{=}1$ it equals documentation-only. Settings frozen at $k{=}5$ (dotted line); three seeds.}
\label{fig:budget}
\end{figure}

Figure~\ref{fig:budget} repeats the main comparisons with 10 and 30 candidates. The router keeps its $k{=}5$ share of usage slots: one in five on StableToolBench and ToolRet (6 of 30 at $k{=}30$) and two in five on TOUCAN (12 of 30).
The held-out lead over the router shrinks on StableToolBench and ToolRet and stays near 9 points on TOUCAN; we suspect part of it comes from the router's growing number of usage slots, which we did not test directly.
The router's lead on fully covered queries persists on ToolRet and TOUCAN at 5, 10 and 30 candidates, and on StableToolBench no difference is detected at 5, 10 and 30.
\syn{}'s gain over documentation on covered queries stays positive at 5, 10 and 30 candidates, with the StableToolBench interval at 30 touching zero.

\section{Appendix D: Tool renaming}

Tool names could let usage text match a query through a shared literal name.
On StableToolBench we mapped every tool identifier to an opaque name and masked literal name mentions in documentation, usage and queries (3,220 documentation mentions, six query mentions and 610 to 723 usage mentions per seed), keeping all settings frozen.
Bare alphabetic or numeric names are masked only in explicit naming contexts, so unquoted brand names can remain.
\syn{} still beats the router on held-out queries by 10.97 points ([6.67, 15.57]), and the labeled difference stays uncertain ($-1.79$, [$-6.84$, 3.94]).

\section{Appendix E: Same-length control}

\begin{figure}[!htbp]
\centering
\includegraphics[width=\columnwidth]{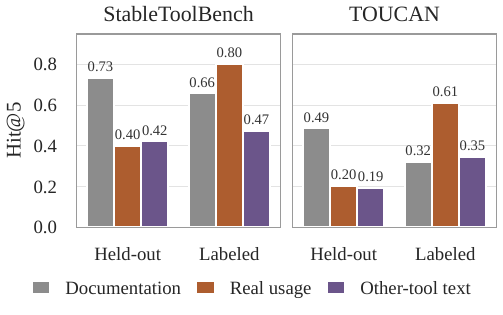}
\caption{Other tools' usage text of the same length gives held-out Hit@5 not detectably different from real usage (StableToolBench $+2.3$, TOUCAN $-0.9$ points) and much lower labeled Hit@5. Usage entries are replaced by length-matched text from other tools' usage; Hit@5, Jina, three seeds.}
\label{fig:length}
\end{figure}

Each usage entry keeps its tool, catalog fields and entry count, and its usage request is replaced by text from other tools in the same pool with exactly the same untruncated Jina-tokenizer length.
Figure~\ref{fig:length} and Table~\ref{tab:length} give the results. All 113,110 StableToolBench entries (the usage entries of the three seed pools, 37,414, 38,008 and 37,688, each drawn from the 90,704 filtered source records) and all 6,617 TOUCAN entries pass this check.
On StableToolBench, 704 entries (0.62\%) have a donor whose gold labels include the target and 411 (0.36\%) keep identical text; on TOUCAN the counts are 408 (6.17\%) and 391 (5.91\%).
Three TOUCAN entries per seed exceed the embedding model's 8,192-token limit and are truncated the same way in the real and control conditions.

\section{Appendix F: Exact values behind the figures}

Tables~\ref{tab:main}, \ref{tab:sweep}, \ref{tab:length}, \ref{tab:agents} and \ref{tab:budget} give the values plotted in Figures~\ref{fig:main}, \ref{fig:rules}, \ref{fig:length}, \ref{fig:agents} and \ref{fig:budget}; Table~\ref{tab:rq1} and Table~\ref{tab:register} give the StableToolBench values and the paired differences behind Figure~\ref{fig:register}, and Table~\ref{tab:m4} the means behind Figure~\ref{fig:m4}.

\begin{table}[h]
\centering
\small
\setlength{\tabcolsep}{3.2pt}
\caption{Values for Figure~\ref{fig:main}: Hit@5 on held-out-test (H) and labeled-test (L) queries, three seeds, frozen settings.}
\label{tab:main}
\begin{tabular}{lcccccc}
\toprule
 & \multicolumn{2}{c}{StableToolBench} & \multicolumn{2}{c}{ToolRet} & \multicolumn{2}{c}{TOUCAN} \\
Rule & H & L & H & L & H & L \\
\midrule
Documentation & .733 & .658 & .306 & .331 & .486 & .323 \\
Folded & .400 & .801 & .290 & .374 & .203 & .612 \\
Untyped grouped & .203 & .745 & .197 & .381 & -- & -- \\
Router & .683 & .826 & .267 & .386 & .419 & .580 \\
\syn{} & .785 & .805 & .294 & .363 & .506 & .488 \\
\bottomrule
\end{tabular}
\end{table}

\begin{table}[h]
\centering
\scriptsize
\setlength{\tabcolsep}{2pt}
\caption{Values for Figure~\ref{fig:rules}: Hit@5 on held-out-test (H) and labeled-test (L) queries for every \syn{} usage weight $w$ and every router setting $d$ (documentation slots out of five), Jina, three seeds. $w{=}0$ is documentation-only. $^{*}$: frozen setting. All other points were computed after the frozen run.}
\label{tab:sweep}
\begin{tabular}{lcccccc}
\toprule
 & \multicolumn{2}{c}{StableToolBench} & \multicolumn{2}{c}{ToolRet} & \multicolumn{2}{c}{TOUCAN} \\
Setting & H & L & H & L & H & L \\
\midrule
\syn{} $w{=}0.0$ & .733 & .658 & .306 & .331 & .486 & .323 \\
\syn{} $w{=}0.1$ & .745 & .707 & .305 & .332 & .501 & .341 \\
\syn{} $w{=}0.2$ & .756 & .749 & .306 & .332 & .511 & .366 \\
\syn{} $w{=}0.3$ & .764 & .785 & .306 & .335 & .519 & .389 \\
\syn{} $w{=}0.4$ & .778 & .805 & .306 & .331 & .526 & .415 \\
\syn{} $w{=}0.5$ & .785$^{*}$ & .805 & .306 & .342 & .527 & .438 \\
\syn{} $w{=}0.6$ & .787 & .794 & .306 & .344 & .524 & .454 \\
\syn{} $w{=}0.7$ & .772 & .799 & .306 & .357 & .519 & .468 \\
\syn{} $w{=}0.8$ & .766 & .796 & .303 & .359 & .516 & .482 \\
\syn{} $w{=}0.9$ & .758 & .783 & .294$^{*}$ & .363 & .506$^{*}$ & .488 \\
\syn{} $w{=}1.0$ & .733 & .756 & .292 & .370 & .496 & .498 \\
\midrule
Router $d{=}1$ & .453 & .803 & .133 & .316 & .246 & .650 \\
Router $d{=}2$ & .584 & .819 & .200 & .369 & .362 & .623 \\
Router $d{=}3$ & .634 & .821 & .224 & .386 & .419$^{*}$ & .580 \\
Router $d{=}4$ & .683$^{*}$ & .826 & .267$^{*}$ & .386 & .455 & .506 \\
\bottomrule
\end{tabular}
\end{table}

\begin{table}[h]
\centering
\caption{Values for Figure~\ref{fig:length} (StableTB: StableToolBench; H, L: held-out and labeled test queries). Docs: documentation only; Real: real usage (folded); Other: other tools' usage of the same length. Last column: Other minus Real, in points, with paired tool-clustered 95\% interval.}
\label{tab:length}
\footnotesize
\setlength{\tabcolsep}{1.6pt}
\begin{tabular}{llcccc}
\toprule
Catalog & & Docs & Real & Other & Difference \\
\midrule
StableTB & H & .733 & .400 & .422 & $+2.28$ [$-0.20$, 5.29] \\
 & L & .658 & .801 & .474 & $-32.66$ [$-40.04$, $-24.93$] \\
TOUCAN & H & .486 & .203 & .194 & $-0.87$ [$-2.06$, 0.24] \\
 & L & .323 & .612 & .347 & $-26.51$ [$-30.76$, $-22.43$] \\
\bottomrule
\end{tabular}
\end{table}

\begin{table}[h]
\centering
\small
\caption{Values for Figure~\ref{fig:agents}: StableToolBench Hit@5. Local: each agent retrieves with its own usage, averaged over agents; Pooled: with all agents' usage. H, L: held-out and labeled test queries.}
\label{tab:agents}
\begin{tabular}{rcccc}
\toprule
Agents & Local H & Local L & Pooled H & Pooled L \\
\midrule
1 & .785 & .805 & .785 & .805 \\
2 & .780 & .781 & .785 & .805 \\
5 & .767 & .713 & .785 & .805 \\
10 & .760 & .674 & .785 & .805 \\
\bottomrule
\end{tabular}
\end{table}

\makeatletter\setlength{\@dblfptop}{0pt}\makeatother
\begin{table*}[t]
\centering
\small
\caption{Values for Figure~\ref{fig:budget} at $k\in\{5,10,30\}$: \syn{} minus the router and minus documentation-only, in Hit@$k$ points with paired tool-clustered 95\% intervals. H: held-out-test; C: labeled-test queries whose every gold tool has usage.}
\label{tab:budget}
\setlength{\tabcolsep}{3pt}
\begin{tabular}{llcccc}
\toprule
Catalog & $k$ & vs router, H & vs router, C & vs docs, H & vs docs, C \\
\midrule
StableToolBench & 5 & $+10.14$ [5.68, 15.05] & $-2.79$ [$-6.82$, 0.66] & $+5.18$ [2.31, 8.39] & $+15.99$ [5.75, 24.12] \\
 & 10 & $+4.55$ [1.83, 7.71] & $+1.27$ [$-3.37$, 5.07] & $+3.93$ [1.35, 6.90] & $+13.96$ [3.93, 22.77] \\
 & 30 & $+6.00$ [2.48, 10.04] & $-0.51$ [$-4.14$, 2.40] & $+4.14$ [1.36, 7.26] & $+8.12$ [0.00, 15.40] \\
ToolRet & 5 & $+2.75$ [0.67, 5.06] & $-17.15$ [$-27.45$, $-8.73$] & $-1.18$ [$-2.85$, 0.24] & $+16.70$ [4.72, 28.64] \\
 & 10 & $+2.09$ [0.52, 4.02] & $-21.16$ [$-29.82$, $-13.15$] & $-0.26$ [$-0.82$, 0.00] & $+19.38$ [7.64, 31.27] \\
 & 30 & $+0.92$ [0.13, 2.05] & $-19.82$ [$-28.57$, $-12.41$] & $-0.65$ [$-1.52$, 0.00] & $+20.71$ [7.18, 34.23] \\
TOUCAN & 5 & $+8.71$ [5.15, 12.39] & $-9.74$ [$-13.40$, $-6.28$] & $+1.95$ [$-0.39$, 4.52] & $+16.77$ [7.53, 25.68] \\
 & 10 & $+9.15$ [5.76, 12.58] & $-11.35$ [$-14.50$, $-8.25$] & $+2.12$ [$-0.26$, 4.70] & $+18.43$ [9.04, 27.58] \\
 & 30 & $+9.36$ [5.35, 15.12] & $-9.49$ [$-11.78$, $-7.34$] & $+2.73$ [0.99, 5.70] & $+19.89$ [9.73, 29.80] \\
\bottomrule
\end{tabular}
\end{table*}

\section{Appendix G: Second embedder (Nomic Embed v1.5)}

Table~\ref{tab:nomic} gives Hit@5 with Nomic Embed v1.5 (136.7M parameters, 8,192-token context) at the settings selected on development tools for this embedder.
Selection used the same 11 weights and four quotas, with the objective of mean held-out and labeled development Hit@5 over three seeds; development and test gold tools are disjoint.
On StableToolBench the selected settings equal Jina's.
Carrying Jina's settings over unchanged gives the same router results and \syn{} results within 1.3 points of these (point estimates).
The Nomic same-length control matches lengths with Nomic's own tokenizer, over the full serialized input including its task prefix; in this StableToolBench control, all 113,110 entries match and none is truncated (longest 538 tokens).
The audit of Nomic's public training data for exposure to these benchmarks is not complete.

\begin{table}[h]
\centering
\small
\setlength{\tabcolsep}{3pt}
\caption{Nomic Embed v1.5, Hit@5 on held-out-test (H), labeled-test (L) and fully covered labeled (C) queries, three seeds, settings reselected on development tools. StableTB: StableToolBench. Same-length: the same-length control, run on StableToolBench only.}
\label{tab:nomic}
\begin{tabular}{llccc}
\toprule
Catalog & Rule & H & L & C \\
\midrule
StableTB & Documentation & .783 & .819 & .812 \\
 & Folded & .621 & .902 & .916 \\
 & Router & .745 & .895 & .901 \\
 & \syn{} & .822 & .879 & .873 \\
 & Untyped grouped & .437 & .841 & .881 \\
 & Same-length & .640 & .774 & .774 \\
ToolRet & Documentation & .388 & .387 & .189 \\
 & Folded & .370 & .434 & .434 \\
 & Router & .353 & .423 & .403 \\
 & \syn{} & .384 & .414 & .298 \\
 & Untyped grouped & .346 & .428 & .474 \\
TOUCAN & Documentation & .679 & .490 & .480 \\
 & Folded & .453 & .727 & .727 \\
 & Router & .603 & .649 & .646 \\
 & \syn{} & .714 & .503 & .493 \\
 & Untyped grouped & .324 & .703 & .704 \\
\bottomrule
\end{tabular}
\end{table}

\section{Appendix H: Workload break-even and a mechanism diagnostic}

\paragraph{Break-even share.}
On StableToolBench, \syn{} gains 4.30 points of Hit@5 over the router ([1.29, 7.45]) at the observed mix of 161 held-out and 149 labeled test queries (48.1\% labeled).
\syn{} is better when fewer than 83.4\% of queries go to labeled tools (interval [59.2\%, 100\%]).
If a labeled query must have usage for every gold tool, the cutoff is 78.4\% ([55.3\%, 100\%]).
An upper endpoint of 100\% includes bootstrap draws with no crossing, so above about 59\% the data do not decide.
These cutoffs assume that the Hit@5 on each kind of query (held-out or labeled) carries over to the target workload, and we did not compute them for ToolRet or TOUCAN.
On StableToolBench, held-out and labeled queries also come from different benchmark groups, so the cutoff mixes tool status with query source.

\paragraph{Usage separates relevant from irrelevant labeled tools.}
Over 340 labeled StableToolBench queries, development queries included, the usage-minus-documentation score separates gold tools from distractor labeled tools at the same documentation rank with an area under the receiver operating characteristic curve (AUC) of 0.817 ([0.780, 0.850]).
This is consistent with usage pushing unrelated labeled tools down, but it is a pooled diagnostic, not a test-only result.

\section{Appendix I: One-entry folding, offset and permuted-usage controls}

All three controls reuse the same embeddings, splits and three seeds, and reproduce every baseline candidate list exactly before scoring.
One-entry folding keeps exactly one usage entry per used tool, chosen by a query-independent content hash, plus one documentation entry per tool; tools without usage keep only documentation.
The constant offset adds a constant $\beta$ to the standardized documentation score of every tool with usage, with $\beta \in \{0, \pm 0.25, \pm 0.5, \ldots, \pm 2, \pm 2.5, \pm 3\}$ selected on development data (selected values: StableToolBench $-0.25$ with Jina and $-0.5$ with Nomic; ToolRet $+0.5$ with both; TOUCAN $-0.25$ with Jina and $+0.5$ with Nomic).
The permuted-usage control permutes the complete set of usage texts and vectors across tools so that no tool keeps its own; entry counts and the global length distribution are kept, but per-entry length is not matched.
We report it at each setting's frozen \syn{} weight.
All results are descriptive on test sets seen in earlier versions of this work.

\begin{table*}[t]
\centering
\small
\setlength{\tabcolsep}{4pt}
\caption{Hit@5 on held-out-test (H) and labeled-test (L) queries for one-entry folding and the constant offset, three seeds. Folded (1 entry): one-entry folding, one usage entry per used tool. Offset: documentation score plus a development-tuned constant for tools with usage.}
\label{tab:controls}
\begin{tabular}{llcccccccccc}
\toprule
 & & \multicolumn{2}{c}{Documentation} & \multicolumn{2}{c}{Folded} & \multicolumn{2}{c}{Folded (1 entry)} & \multicolumn{2}{c}{\syn{}} & \multicolumn{2}{c}{Offset} \\
Embedder & Catalog & H & L & H & L & H & L & H & L & H & L \\
\midrule
Jina & StableToolBench & .733 & .658 & .400 & .801 & .553 & .682 & .785 & .805 & .770 & .626 \\
 & ToolRet & .306 & .331 & .290 & .374 & .290 & .360 & .294 & .363 & .282 & .342 \\
 & TOUCAN & .486 & .323 & .203 & .612 & .287 & .503 & .506 & .488 & .509 & .286 \\
Nomic & StableToolBench & .783 & .819 & .621 & .902 & .687 & .839 & .822 & .879 & .826 & .776 \\
 & ToolRet & .388 & .387 & .370 & .434 & .371 & .422 & .384 & .414 & .384 & .406 \\
 & TOUCAN & .679 & .490 & .453 & .727 & .541 & .616 & .714 & .503 & .582 & .552 \\
\bottomrule
\end{tabular}
\end{table*}

\begin{table}[h]
\centering
\small
\setlength{\tabcolsep}{3pt}
\caption{Permuted-usage control: \syn{} with every tool's usage replaced by another tool's, minus \syn{}, at the frozen weight (H, L: held-out and labeled test queries; Emb.: embedding model; StableTB: StableToolBench), in Hit@5 points with paired tool-clustered 95\% intervals.}
\label{tab:othertool}
\scriptsize
\setlength{\tabcolsep}{2pt}
\begin{tabular}{llcc}
\toprule
Emb. & Catalog & H & L \\
\midrule
Jina & StableTB & $+5.38$ [2.31, 9.05] & $-54.59$ [$-61.07$, $-46.88$] \\
 & ToolRet & $+0.92$ [0.14, 1.87] & $-6.18$ [$-9.33$, $-3.55$] \\
 & TOUCAN & $+1.56$ [0.54, 2.68] & $-46.65$ [$-54.49$, $-39.29$] \\
Nomic & StableTB & $+5.80$ [2.12, 10.10] & $-49.89$ [$-56.13$, $-41.95$] \\
 & ToolRet & $+0.00$ [0.00, 0.00] & $-6.68$ [$-9.58$, $-4.23$] \\
 & TOUCAN & $+1.39$ [0.47, 2.65] & $-45.87$ [$-55.16$, $-37.32$] \\
\bottomrule
\end{tabular}
\end{table}

Against documentation-only, one-entry folding lowers held-out Hit@5 by 18.01 ([12.80, 23.33]), 1.57 ([0.38, 3.19]; the same held-out Hit@5 and interval as full folding on ToolRet, consistent with most covered ToolRet tools having a single entry) and 19.98 ([10.93, 28.82]) points with Jina on StableToolBench, ToolRet and TOUCAN, and by 9.52 ([4.60, 15.40]), 1.70 ([0.39, 3.40]) and 13.78 ([7.94, 19.75]) points with Nomic.
The offset minus \syn{} on held-out queries is $-1.45$ ([$-4.25$, 1.34]), $-1.18$ ([$-2.96$, 0.40]) and $+0.35$ ([$-1.49$, 2.01]) with Jina, and $+0.41$ ([$-1.70$, 2.69]), $0.00$ ([$-1.33$, 1.33]) and $-13.18$ ([$-21.68$, $-6.28$]) with Nomic.
Against the router, the offset is higher on held-out queries in five of six settings (by 1.6 to 9.1 points; on TOUCAN with Nomic it is 2.1 lower) and lower on labeled queries in all six (by 1.7 to 29.4 points), point estimates from Table~\ref{tab:controls} and the router rows; with Jina on StableToolBench and TOUCAN its labeled Hit@5 is also below documentation-only.
On labeled queries it is $-17.90$ ([$-26.67$, $-8.19$]), $-2.06$ ([$-4.94$, 0.44]) and $-20.21$ ([$-28.89$, $-11.34$]) with Jina, and $-10.29$ ([$-16.04$, $-2.89$]), $-0.78$ ([$-3.65$, 1.61]) and $+4.96$ ([$-1.88$, 12.32]) with Nomic.

\section{Appendix J: Equal-entry and generated-request controls}

Table~\ref{tab:m29} gives Hit@5 and Table~\ref{tab:m29diff} the paired differences. These StableToolBench controls use Jina embeddings, the fixed splits and three seeds, and replay every earlier candidate list exactly before scoring.
\emph{Raw, all} is the untyped grouped index (one documentation entry per tool plus every raw usage entry); it differs from Table~\ref{tab:main} by one labeled hit because ties are broken by entry order instead of tool identifier.
\emph{Raw, one} keeps one raw usage entry per used tool, chosen by a seeded content hash.
\emph{Appended} indexes concatenate one or all usage examples to the used tool's documentation, so every tool has exactly one entry; with all usage, 15 to 17 tools per seed exceed the encoder limit and are truncated, and with one example none are.
For the generated requests, Llama-3.1-8B-Instruct wrote one request per catalog tool from its public documentation alone, with greedy decoding and at most 96 new tokens; it never saw test queries, usage or the held-out set.
None of the 3,180 requests matches any of the 729 StableToolBench queries exactly after normalization or with Jina cosine $\geq 0.95$.
\emph{Raw + generated} adds a generated request for each tool without usage to Raw, all; \emph{One per tool} has one documentation entry and one request per tool (a real usage example for used tools, a generated request for the rest), 6,360 entries in all; \emph{Generated, all} adds a generated request to every tool as a separate entry next to its documentation entry and uses no real usage.
\emph{Generated, all} is the Re-Invoke-inspired index of the main text: it is inspired by Re-Invoke \citep{chen2024reinvoke} but generates one request per tool instead of ten and omits its intent extraction and multi-view ranking.
All comparisons here were made after the frozen results and are descriptive.

\begin{table}[h]
\centering
\small
\setlength{\tabcolsep}{4pt}
\caption{Equal-entry and generated-request controls: StableToolBench Hit@5 on held-out-test (H) and labeled-test (L) queries, Jina, three seeds.}
\label{tab:m29}
\begin{tabular}{lcc}
\toprule
Index & H & L \\
\midrule
Documentation & .733 & .658 \\
Folded & .400 & .801 \\
Raw, all (untyped grouped) & .203 & .743 \\
Raw, one & .358 & .521 \\
Appended, one example & .429 & .671 \\
Appended, all usage & .424 & .617 \\
Raw + generated & .300 & .747 \\
One per tool & .472 & .530 \\
Generated, all & .665 & .658 \\
Router & .683 & .826 \\
\syn{} & .785 & .805 \\
\bottomrule
\end{tabular}
\end{table}

\begin{table}[h]
\centering
\scriptsize
\setlength{\tabcolsep}{2pt}
\caption{Paired differences for Table~\ref{tab:m29}, in Hit@5 points with tool-clustered 95\% intervals. App.: appended; Raw+gen.: raw usage plus generated requests; Gen., all: the Re-Invoke-inspired index; docs: documentation-only; other names as in Table~\ref{tab:m29}. H, L: held-out and labeled test queries.}
\label{tab:m29diff}
\setlength{\tabcolsep}{1.2pt}
\begin{tabular}{lcc}
\toprule
Comparison & H & L \\
\midrule
App.\ one $-$ docs & $-30.43$ [$-37.50$, $-23.53$] & $+1.34$ [$-5.37$, 7.13] \\
App.\ all $-$ docs & $-30.85$ [$-37.53$, $-24.31$] & $-4.03$ [$-12.38$, 4.88] \\
App.\ all $-$ folded & $+2.48$ [$-1.39$, 6.58] & $-18.34$ [$-27.40$, $-9.07$] \\
App.\ one $-$ raw, one & $+7.04$ [3.21, 11.18] & $+14.99$ [8.15, 21.49] \\
One per tool $-$ docs & $-26.09$ [$-33.74$, $-18.43$] & $-12.75$ [$-21.77$, $-3.07$] \\
One per tool $-$ raw, one & $+11.39$ [6.67, 16.86] & $+0.89$ [0.00, 2.58] \\
Raw+gen.\ $-$ raw, all & $+9.73$ [4.98, 15.22] & $+0.45$ [$-0.86$, 2.21] \\
Gen., all $-$ docs & $-6.83$ [$-13.07$, $-1.26$] & $0.00$ [$-5.16$, 5.09] \\
\syn{} $-$ gen., all & $+12.01$ [6.40, 18.16] & $+14.77$ [4.51, 25.31] \\
\bottomrule
\end{tabular}
\end{table}

\section{Appendix K: Agents}

\paragraph{Heterogeneous agents.}
StableToolBench's usage pool is split among five agents, using the public tool categories.
For $\alpha \in \{0.1, 0.3, 1\}$ and each seed, we draw five-agent shares from Dirichlet$(\alpha,\ldots,\alpha)$ for every distinct category set of source queries, then give each source query one owner by hash sampling from those shares; $\alpha$ is per component, so the total concentration is $5\alpha$. The hash split assigns each source query to an agent by a hash; it is the same partition as the five-agent case of the homogeneous split below, with identical candidate lists.
The two report different gains because Table~\ref{tab:m4} weights each agent's test queries by its own category mix, while Table~\ref{tab:agents} weights all test queries equally (\syn{} pooled minus local: held-out $+0.34$ against $+1.78$ points, labeled $+5.26$ against $+9.26$).
All usage examples from one source query stay with one agent, nothing is replicated, and agent sizes are left unequal.
Each agent $j$'s mix over tool categories $c$, $p_j(c)$, is estimated from its own records only.
A test query spreads one unit evenly over its gold tools' categories, giving $f(q,c)$; with $m_c=\sum_q f(q,c)$ in each test set (held-out or labeled), the agent weights query $q$ by $\sum_c p_j(c)\,f(q,c)/m_c$, normalized within agent, seed and test set, so each category gets the share it has in the agent's records.
No scores or query text enter the weights.
Across all partition settings, agent and seed workloads have Kish effective sample sizes $1/\sum_q \pi_q^2$, with $\pi_q$ the normalized query weights, of 5.9 to 62.9 (held-out) and 14.6 to 82.3 (labeled). Single-agent extremes are exploratory: the smallest held-out workload has a Kish effective sample size of 5.9 queries.
Category mass with no test query in the test set is 6.5 to 64.4\% (held-out) and 0.7 to 26.2\% (labeled), and weights are renormalized over the supported categories.
At $\alpha{=}0.3$, the workload-weighted share of labeled queries whose every gold tool has usage is 63.8 to 85.3\% locally and 80.9 to 97.4\% pooled; held-out coverage is zero throughout.
Besides retrieving with its own records or with all agents' records, an agent can pool with itself and the two agents whose category mixes are closest in Jensen--Shannon divergence, chosen without outcomes.
Every rule keeps its frozen setting for every agent, split and sharing condition.
Table~\ref{tab:m4} reports pooling with all agents; with the two closest agents, folding's average held-out loss is smaller (6.76 to 11.09 points, every interval below zero) and \syn{}'s average held-out change again stays within 1.4 points of zero.
The worst-agent value takes the lowest agent change in each seed and averages it over seeds; the share losing is the fraction of agents with a negative change, averaged over seeds. The bootstrap recomputes both in every draw, keeping seeds and agents together.
Documentation-only lists do not change with pooling, nor does the router's held-out Hit@5, because held-out tools never fill a usage slot.
A pooled list equals that of one agent holding the union of records, checked on every list.

\begin{table*}[t]
\centering
\small
\setlength{\tabcolsep}{4pt}
\caption{Pooling with all agents minus retrieving locally, StableToolBench, Jina, five agents, three seeds: mean change in Hit@5 points with tool-clustered 95\% interval, worst agent, and share of agent-seed cells that lose. H, L: held-out and labeled test queries.}
\label{tab:m4}
\begin{tabular}{llccccc}
\toprule
Split & Rule & H mean & H worst agent & H share losing & L mean & L worst agent \\
\midrule
$\alpha{=}0.1$ & Folded & $-$13.25 [$-$22.13, $-$5.29] & $-$31.20 & 1.00 & 2.89 [0.16, 5.98] & $-$2.63 \\
 & Router & 0.00 [0.00, 0.00] & 0.00 & 0.00 & 3.02 [0.72, 5.88] & $-$3.08 \\
 & \syn{} & $-$1.32 [$-$5.99, 2.98] & $-$9.93 & 0.40 & 7.25 [4.31, 10.70] & 3.64 \\
\midrule
$\alpha{=}0.3$ & Folded & $-$12.10 [$-$19.04, $-$5.93] & $-$20.09 & 1.00 & 4.56 [2.40, 7.11] & 1.19 \\
 & Router & 0.00 [0.00, 0.00] & 0.00 & 0.00 & 4.15 [1.72, 7.01] & $-$1.06 \\
 & \syn{} & $-$0.86 [$-$4.58, 2.05] & $-$5.70 & 0.47 & 6.08 [3.67, 8.93] & 2.51 \\
\midrule
$\alpha{=}1$ & Folded & $-$13.20 [$-$21.34, $-$6.07] & $-$16.75 & 1.00 & 3.45 [0.62, 6.42] & $-$1.73 \\
 & Router & 0.00 [0.00, 0.00] & 0.00 & 0.00 & 5.99 [3.52, 8.88] & 3.64 \\
 & \syn{} & 0.97 [$-$0.88, 2.75] & $-$0.37 & 0.20 & 5.66 [2.90, 8.66] & 2.57 \\
\midrule
Hash & Folded & $-$12.91 [$-$20.90, $-$5.87] & $-$14.52 & 1.00 & 4.93 [1.15, 8.77] & 2.08 \\
 & Router & 0.00 [0.00, 0.00] & 0.00 & 0.00 & 7.05 [3.86, 10.72] & 4.87 \\
 & \syn{} & 0.34 [$-$1.53, 1.77] & $-$1.08 & 0.33 & 5.26 [2.28, 8.29] & 1.50 \\
\bottomrule
\end{tabular}
\end{table*}

\paragraph{Homogeneous split.}
We also split the pool among 1, 2, 5 or 10 agents by hashing each source query and score with the frozen \syn{} weight (Figure~\ref{fig:agents}, Table~\ref{tab:agents}).
At ten agents, pooling beats the average single agent by 2.48 held-out points ([0.64, 4.51]) and 13.18 labeled points ([6.04, 18.70]).
The gain measures what each agent lacks on its own; the experiment tests no benefit specific to federation.

\section{Appendix L: Rank fusion, thresholded router and text-type control}

\paragraph{Rank fusion and thresholded router.}
Each tool has a documentation rank $r_{\mathrm{doc}}$ over the whole catalog and, if it has usage, a usage rank $r_{\mathrm{use}}$ among tools with usage, where its usage score is its best-matching usage entry.
With usage weight $\omega$ and rank constant $\kappa$, additive RRF scores a tool by $(1-\omega)/(\kappa+r_{\mathrm{doc}}) + \omega/(\kappa+r_{\mathrm{use}})$ and gives a tool without usage only the first term.
The renormalized variant scores a tool without usage by $1/(\kappa+r_{\mathrm{doc}})$, so it is not penalized for lacking usage.
The best-field variant takes the larger of the two weighted terms.
An earlier screen on development data searched one setting shared across catalogs ($\omega \in \{0.3,\ldots,0.7\}$, $\kappa \in \{10, 30, 60\}$, 45 settings) and found none that met its preset development Hit@5 target on all three catalogs, so it reported no test scores.
For the results here, we selected $\omega$ and $\kappa$ separately per catalog and rule on development data (equal weight on held-out and labeled development Hit@5) and then scored the test sets once.
Selected settings: additive RRF $\omega{=}0.3$, $\kappa{=}10$ on StableToolBench and ToolRet and $\omega{=}0.4$, $\kappa{=}10$ on TOUCAN; renormalized $\omega{=}0.5$, $\kappa{=}30$ on StableToolBench and $\omega{=}0.7$, $\kappa{=}10$ on ToolRet and TOUCAN; best-field $\omega{=}0.3$, $\kappa{=}10$ on StableToolBench and ToolRet and $\omega{=}0.5$, $\kappa{=}10$ on TOUCAN.
The thresholded router gives its first slot to the best usage match only when that match's cosine reaches a threshold set for a 5\% false-positive rate on held-out development queries, and fills the other slots from documentation.
On StableToolBench and ToolRet the selected best-field setting never lets usage win and reproduces documentation-only.
The test sets were seen in earlier versions of this work, so these comparisons are descriptive; Table~\ref{tab:rrfdiff} gives the paired differences.

\begin{table}[h]
\centering
\scriptsize
\setlength{\tabcolsep}{1pt}
\caption{\syn{} minus each fusion baseline, in Hit@5 points with tool-clustered 95\% intervals. H, L: held-out and labeled test queries; StableTB: StableToolBench.}
\label{tab:rrfdiff}
\begin{tabular}{llcc}
\toprule
Catalog & Baseline & H & L \\
\midrule
StableTB & Additive RRF & $+15.73$ [10.53, 21.19] & $+6.26$ [0.00, 12.11] \\
 & Renormalized RRF & $+0.62$ [$-0.22$, 1.57] & $+3.80$ [0.43, 7.60] \\
 & Best-field RRF & $+5.18$ [2.31, 8.39] & $+14.77$ [5.51, 22.55] \\
 & Thresholded router & $+5.38$ [2.43, 8.71] & $+6.94$ [1.15, 11.75] \\
ToolRet & Additive RRF & $+0.26$ [$-0.93$, 1.50] & $+1.24$ [$-1.17$, 4.11] \\
 & Renormalized RRF & $-0.26$ [$-1.61$, 1.02] & $+0.14$ [$-1.80$, 2.55] \\
 & Best-field RRF & $-1.18$ [$-2.85$, 0.24] & $+3.16$ [0.60, 6.18] \\
 & Thresholded router & $-0.78$ [$-2.20$, 0.40] & $+1.42$ [$-0.64$, 3.84] \\
TOUCAN & Additive RRF & $+15.99$ [8.38, 23.40] & $-4.42$ [$-13.23$, 3.63] \\
 & Renormalized RRF & $+2.77$ [0.61, 4.98] & $-3.76$ [$-8.79$, 0.96] \\
 & Best-field RRF & $+10.53$ [6.52, 14.70] & $-11.84$ [$-14.98$, $-9.02$] \\
 & Thresholded router & $+1.95$ [$-0.39$, 4.52] & $+14.46$ [5.70, 22.95] \\
\bottomrule
\end{tabular}
\end{table}

\paragraph{Text-type control.}
For each usage entry we pick a donor: a different tool with public documentation, taken from a usage record whose source query does not list the recipient tool among its gold tools and whose request text is not identical to the recipient's.
A seeded content hash orders the scan over donors, and no query or outcome chooses them.
The entry's text is then replaced either by the donor's request text or by the donor's documentation, each matched exactly to the original entry's full token length; short texts are repeated to reach the length.
The entry keeps its own tool, catalog fields and position, and folding is otherwise unchanged.
Table~\ref{tab:register} gives the paired differences. On StableToolBench about 49\% of request replacements and 62\% of documentation replacements needed repetition; on TOUCAN the shares are about 49\% and 7\%, and three entries per seed exceed the encoder limit with both replacements.

\begin{table}[h]
\centering
\scriptsize
\setlength{\tabcolsep}{0.8pt}
\caption{Text-type control, in Hit@5 points with tool-clustered 95\% intervals. D: other tools' documentation; R: other tools' requests; docs only: the documentation-only baseline. No-rep.: only entries where neither replacement was repeated, which also changes which tools keep usage. StableTB: StableToolBench.}
\label{tab:register}
\begin{tabular}{llcc}
\toprule
Catalog & Comparison & H & L \\
\midrule
StableTB & D $-$ R & $+21.53$ [15.04, 28.29] & $+14.32$ [8.18, 21.14] \\
 & R $-$ folded & $+2.48$ [0.21, 5.13] & $-29.08$ [$-35.66$, $-20.87$] \\
 & D $-$ docs only & $-9.32$ [$-13.68$, $-5.32$] & $-0.45$ [$-4.05$, 2.58] \\
 & No-rep.\ D$-$R & $+12.42$ [7.25, 18.13] & $+5.15$ [0.00, 11.99] \\
TOUCAN & D $-$ R & $+22.97$ [12.54, 32.88] & $+6.80$ [3.24, 11.25] \\
 & R $-$ folded & $+0.61$ [$-2.00$, 2.59] & $-30.13$ [$-35.08$, $-25.43$] \\
 & D $-$ docs only & $-4.77$ [$-6.88$, $-2.65$] & $+5.56$ [3.35, 7.80] \\
 & No-rep.\ D$-$R & $+16.86$ [8.96, 24.45] & $+0.40$ [$-4.02$, 5.05] \\
\bottomrule
\end{tabular}
\end{table}

\section{Appendix M: Held-out-share sweep}

On TOUCAN we held out tools whose name hash falls below 1, 3 or 5 in ten buckets, which gives 200, 627 and 1,049 of 2,159 tools; the sets are nested, and the 30\% set is the one used everywhere else.
The development and test buckets are unchanged.
For each share we rebuilt the usage pool from the same candidate records with the same exclusions (any record that names a held-out tool as gold or mentions its name, and near-duplicates of evaluation queries), the same hash-chosen 80\% keep and the same cap of eight entries per tool; the 30\% pool is identical to the one used everywhere else.
Our primary query set is the 123 held-out test queries at the 10\% share (21 tool clusters), which are held out at every share.
In the matched-volume variant we subsample source records at the 10\% and 30\% shares to the record count at 50\%, keeping all usage examples of a selected source query together.
New usage at the 10\% share needed 2,455 usage entries that had not been embedded before; we embedded them with the same Jina model and recorded this change to the protocol before scoring.
All rules keep the frozen TOUCAN settings (\syn{} weight $w{=}0.9$, router documentation slots $d{=}3$); Table~\ref{tab:share} gives the paired differences.
On each share's own full held-out set, folding gives 0.222, 0.203 and 0.208 against documentation-only 0.455, 0.486 and 0.395, and \syn{} gives 0.442, 0.506 and 0.416.

\begin{table}[h]
\centering
\scriptsize
\setlength{\tabcolsep}{2pt}
\caption{Held-out-share sweep on the fixed 123 queries, in Hit@5 points with tool-clustered 95\% intervals. Usage ``all'' keeps every remaining record; ``matched'' subsamples records to the count at the 50\% share. Docs: documentation-only.}
\label{tab:share}
\begin{tabular}{llcc}
\toprule
Share & Usage & Folded $-$ docs & \syn{} $-$ folded \\
\midrule
10\% & all & $-23.31$ [$-39.93$, $-13.82$] & $+21.95$ [9.79, 42.65] \\
 & matched & $-19.78$ [$-34.10$, $-11.78$] & $+20.33$ [8.69, 39.58] \\
30\% & all & $-20.87$ [$-36.82$, $-12.15$] & $+19.51$ [10.84, 35.29] \\
 & matched & $-20.33$ [$-35.78$, $-11.85$] & $+20.60$ [11.39, 36.81] \\
50\% & all & $-15.72$ [$-28.02$, $-8.55$] & $+16.80$ [8.91, 30.08] \\
\bottomrule
\end{tabular}
\end{table}

\end{document}